\documentclass[12pt,preprint]{emulateapj}
\begin{document}

\newcommand{\etal}{{\it et al.}}
\newcommand{\ls}{{\it CFHTLS}}

\slugcomment{Accepted for publication in ApJ}

\title{The Size Function of Galaxy Disks out to $z \sim 1$ from the Canada-France-Hawaii-Telescope Legacy Survey}

\author{Anudeep Kanwar\altaffilmark{1,2}, Luc Simard \altaffilmark{2}, David Schade\altaffilmark{2}, \& Stephen D.J. Gwyn\altaffilmark{1,2}}

\altaffiltext{1}{Department of Physics and Astronomy, University of Victoria, P.O. Box 3055, STN CSC, Victoria BC, V8W 3P6, Canada}
\altaffiltext{2}{National Research Council of Canada, Herzberg Institute of Astrophysics, 5071 West Saanich Road, Victoria BC, V9E 2E7, Canada}
\altaffiltext{*}{Based on observations obtained with MegaPrime/MegaCam, a joint project of CFHT and CEA/DAPNIA, at the Canada-France-Hawaii Telescope (CFHT) which is operated by the National Research Council (NRC) of Canada, the Institut National des Science de l'Univers of the Centre National de la Recherche Scientifique (CNRS) of France, and the University of Hawaii. This work is based in part on data products produced at TERAPIX and the Canadian Astronomy Data Centre as part of the Canada-France-Hawaii Telescope Legacy Survey, a collaborative project of NRC and CNRS.}

\begin{abstract}
The formation and growth of galaxy disks over cosmic time is crucial to our understanding of galaxy formation. Despite steady improvements in the size and quality of disk samples over the last decade, many aspects of galaxy disk evolution remain unclear. Using two square degrees of deep, wide-field $i^{\prime}$-band imaging from the Canada-France-Hawaii Telescope Legacy Survey, we compute size functions for 6000 disks from $z=0.2$ to $z=1$ and explore luminosity and number density evolution scenarios with an emphasis on the importance of selection effects on the interpretation of the data. We also compute the size function of a very large sample of disks from the Sloan Digital Sky Survey to use as a local ($z \simeq$ 0.1) comparison. CFHTLS size functions computed with the same fixed luminosity-size selection window at all redshifts exhibit evolution that appears to be best modelled by a pure number density evolution. The $z = 0.3$ size function is an excellent match to the $z = 0.9$ one if disks at the highest redshift are a factor of 2.5 more abundant than in the local universe. The SDSS size function would also match the $z = 0.9$ CFHTLS size function very well with a similar change in number density. On the other hand, the CFHTLS size functions computed with a varying luminosity-size selection window with redshift remain constant if the selection window is shifted by 1.0$-$1.5 mag towards fainter magnitudes with decreasing redshift. There is a weak dependence on disk scale length with smaller ($h \lesssim 4$ kpc) disks requiring more luminosity evolution than larger ones. Given that changes in number density are primarily due to mergers and that current estimates of merger rates below $z$ = 1 are low, luminosity evolution appears to be a more plausible scenario to explain the observations.

\end{abstract}
\keywords{galaxies: formation, evolution}

\section{Introduction}

The size evolution of disk galaxies provides a unique constraint for galaxy formation models. 
In the $\Lambda$CDM paradigm, structure forms hierarchically. Primordial density perturbations gradually attract dark matter into haloes that become the sites of galaxy formation. Through tidal torques, these haloes and their gas contents acquire angular momentum. The gas within the halo cools and condenses to form a flattened disk, supporting itself through rotation \citep{fall}. The first disks formed in these haloes are small and dense. After several mergers as well as gas accretion, these disks grow to become giant spirals and subsquently, ellipticals \citep{steinnav,toomres}. Mergers are thus an important mechanism in the evolution of galaxies. They are also quite violent processes. Minor mergers are thought to lead to the growth of a bulge \citep{barnes} and to an increase in disk thickness through heating \citep{toth}. Equal mass mergers can completely change the original morphology of a galaxy by turning disks into spheroids. Although the gas content of the constituent galaxies may permit a re-growth of a disk around a bulge galaxy \citep{robertson}, a pure, bulgeless disk galaxy grows mainly through quiescent accretion.  The cold, fragile structure of disk galaxies thus make them unique probes of hierarchical mass assembly; they provide a benchmark for the sizes of galaxies at any epoch as they give the sizes of objects that have been relatively undisturbed through cosmic time.

Generally, models predict that large disk galaxies should only be stable at recent ($z<1$) epochs \citep{momao,vandenB}. However,  previous surveys \citep[e.g.,][]{lilly98, roche98} have found that the size distribution of disks remained relatively unchanged at $z<1$ once about 1 mag of luminosity evolution had been taken into account. This suggests that size evolution and the bulk of mass assembly must have occurred \textit{prior} to $z=1$. Although luminosity evolution is a natural consequence of an aging stellar population \citep{tinsley}, previous surveys have also found a range of values; from little or no evolution \citep{simard99, ravi} to as much as 1 magnitude or more \citep{schade96a,lilly98,barden,sargent}. One critical aspect of these surveys is the treatment of selection effects as different treatments can lead to different conclusions from the same dataset. 
Despite concluding otherwise, both \citet{simard99} and \citet{ravi} were able to produce a 1 magnitude evolution with their respective samples by redefining their selection criteria.
 This is a systematic problem that is not alleviated with larger samples. For example, one may use a fixed surface brightness survey window with redshift and follow changes in galaxy properties within this window. Alternatively, one may attempt to track the same galaxy population across all redshifts by sliding the survey window according to some assumptions on the evolution of this population. 

Earlier disk galaxy surveys did not have the advantage of the wide imaging capabilities afforded by recent studies. These earlier observations had to deal with the additional complexity of cosmic variance due to sampling in small fields. As survey field sizes have grown larger, sample sizes have also grown substantially. As a result, more statistically significant samples are now available, and the expanded sky coverage of these surveys minimizes the effect of cosmic variance. Over a decade ago, \citet{vogt} used a sample of 16 galaxies and found that large, massive disks were in place by $z\approx 1$.  \citet{roche98} used a more sizeable sample of 347 objects and found that disks have undergone a surface brightness evolution of approximately 0.95 mag over $0.2<z<0.9$ through a combination of size and luminosity evolution with most of the size evolution occurring prior to $z=1$. 
Using 341 objects selected from the Canada-France-Hawaii Redshift survey, \citet{lilly98} found that the size function of larger disks (disk scale length $h >$ 3.2$h^{-1}$ kiloparsecs) is approximately constant to $z\approx 1$ and that disks are approximately 0.8 mag brighter at $z=0.7$. These studies seem to support a model where large disk galaxies are in place by $z=1$ in the same number density and physical sizes as they are today. In addition, they claim that the luminosity evolution is consistent with a simple, passively aging population. However, other studies find that the luminosity evolution in disks is minimal over this time. With a sample of 190 field galaxies from the Deep Extragalactic Evolutionary Probe (DEEP), \cite{simard99} performed a careful analysis of selection effects and found that there has been no evolution in disk surface brightness. \citet{ravi} used 1508 galaxies from the Great Observatories Origins Deep Survey (GOODS) and also found that there has been little or no evolution ($\leq$ 0.4 mag) in the surface brightness of disks over the redshift range $0.2<z<1.25$.  
More recent studies came to the conclusion that disk-dominated galaxies actually evolved in luminosity. \citet{barden} used 5664 disk-dominated galaxies from the Galaxy Evolution from Morphologies and SEDs (GEMS) survey and found that there has been a brightening of approximately 1 magnitude to $z\le 1.1$. By applying the same amount of surface brightness evolution as measured by \citet{barden} to a sample of 36 disks in the Hubble Ultra Deep Field, \citet{truj2005} find that these disks have had moderate ($\sim$ 25$\%$) inside-out growth since $z = 1.1$. As part of the COSMOS project, \citet{sargent} examined approximately 12,000 disk-dominated galaxies and found that the number density of large disks has remained fairly constant out to $z=1$.  Of these however, they found that their very largest disks ($h >$ 10 kpc) are only 60\% as abundant as today, but inferred that some of these objects may have a more significant bulge component in the present day.  This could be interpreted as a transformation from bulgeless disks at $z=1$ into relatives of the Milky Way. Studies at high redshift now benefit from a better understanding of local galaxy populations thanks to the Sloan Digital Sky Survey \citep{strauss}, but a clear consensus on the evolution of disk galaxies since $z = 1$ has yet to emerge. 
This paper focuses on the size evolution of galaxy disks out to $z \sim 1$ from the deep, wide-field imaging of the Canada-France-Hawaii Telescope Legacy Survey (CFHTLS). It is organized as follows. The data and bulge+disk decompositions are described in Sections~\ref{imaging_data} and~\ref{methodology}. The sample selection and selection effects are discussed in Section~\ref{sampsel}. In particular, we will explore how different treatments of selection effects may be interpreted as different evolutionary models.  The disk size functions are presented in Sections~\ref{the_sizefunc} and~\ref{results}. The implications of our results for the evolution of galaxy disks are discussed in Section~\ref{discussion}. The cosmology adopted throughout this paper is ($H_0, \Omega_{m}, \Omega_{\Lambda}$) = (70, 0.3 , 0.7).

\section{Imaging Data}\label{imaging_data}
The images used for this study were obtained as part of the Deep component of the Canada-France-Hawaii Telescope Legacy Survey (CFHTLS-Deep). The CFHTLS is a large five-year project that began in 2003. The overall goal of CFHTLS-Deep  is to obtain a better understanding of the early universe through detection of thousands of supernovae and the study of the galaxy population. Covering nearly four square degrees, the survey uses large statistical samples to build stronger constraints on galaxy evolution and the star formation history of the Universe. Such a large survey area is accessible thanks to the MegaPrime prime-focus mosaic imager (9$\times$4 mosaic of 2080 pixel $\times$ 4622 pixel detectors). With a scale of 0.187''/pixel, the full mosaic has a 0.96 deg $\times$ 0.96 deg field of view.  The morphological analysis was performed using the $i^{\prime}$ band images, and other bandpasses were used for the computation of photometric redshifts. We used two deep fields for this study: D1 ($\alpha$= 02:26:00, $\delta$ =$-$04:30:00) and D3 ($\alpha$=14:17:54,+52:30:31). The D3 field partially overlaps the well-known Groth-Westphal Strip region \citep{rhodes2000}. Analysis for each field was performed on stacked $i^{\prime}$-band images (6580$-$9120\AA) consisting of 150 exposures totaling 12.9 hours and 96 exposures totaling 13.9 hours, respectively. The depths of our D1 and D3 image stacks (5-$\sigma$ detection limit) were 26.5 $\pm$ 0.1 mag and 26.7 $\pm$ 0.1 mag respectively.

All previous studies of disks at high redshift have used space-based imaging from the Hubble Space Telescope. While HST imaging clearly offers better spatial resolutions, ground-based observations with wide-field imagers can deliver images over larger fields with longer integration times. Large field size is important to reduce the effects of cosmic variance which can be significant even at square degree scales \citep{somerville04, trenti}.
Longer integration times mean that lower surface brightness levels can be probed. The limited spatial resolution of ground-based observations is offset by an increase in the radial range over which photometric decompositions can be performed. While it is true that the structural parameters of high-redshift bulges cannot be measured from the ground due to seeing-limited resolution, disk scale length measurements can compare very well with space-based results thanks to larger disk apparent sizes and larger fitting ranges.

\section{Methodology}\label{methodology}
\subsection{Image Processing and Point-Spread-Function}\label{imageproc}
Basic preprocessing (bias removal, flat fielding and fringe correction) was performed using the  Elixir pipeline (Magnier \& Cuillandre 2004), and elixir-processed images were then retrieved from the Canadian Astronomical Data Centre. All the images were astrometrically and photometrically calibrated and then stacked using an early version of the MegaPipe reduction pipeline \citep{gwyn}. Sources are extracted using the \texttt{SExtractor} routine \citep{bertin}, and a catalogue of objects is generated. Basic photometry in all 5 bands was done using Kron-style apertures using the double image mode of \texttt{SExtractor}. Photometric redshifts and rest-frame absolute magnitudes were computed following \citet{gwyn}. 
Photometric redshifts were calculated using a template fitting method. In this method, galaxy broadband colours in Megacam $u^*$, $g^\prime$, $r^\prime$, $i^\prime$ and $z^\prime$ are used to create a low resolution spectral energy distribution (SED) which is then compared to a number of known galaxy template spectra (eg. star-forming, quiescent etc). Here we use template spectra from \citet{cww}. For a single galaxy, a suitable template is determined by the one that returns the lowest $\chi^2$ value.   Using features that are common to both spectra, the galaxy spectrum is shifted until it is best aligned with the template distribution. This shift is used in the standard redshift relation $z=(\lambda-\lambda_0)/\lambda_0$ to determine the redshift. There is not a single ``feature" that is used for $\lambda_0$; the optimal shift is found by an overall agreement between the galaxy SED and the template. The relative error in the photometric redshift measurements is about $\delta z/(1+z)   = 0.11$ \citep{nuij}.

The point spread function of the images was built using the DAOPHOT \citep{stetson} package within the Image Reduction and Analysis Facility (IRAF). A variable Moffat point-spread function (PSF) was constructed for each chip. Typically, 75 point source objects were used to generate the model point spread function for each chip. Selected objects were in focus, had very high \texttt{SExtractor} stellarity indexes and were fairly well isolated with regular, circular isophotes. A model for the point spread function was determined through several iterations. Point sources were convolved with the PSF model and then subtracted from the original image at the locations of the objects used to derive the PSF. The subtracted image was then visually inspected for residuals. A good PSF model  will yield clean image residuals with no systematics present. Possible systematic errors for the point spread function include a very bright or dark spot in the center of the object or dark spots within the PSF fitting radius. Parameters were then adjusted based on the type of systematics present, and the model was recomputed. Once all systematics were minimized, the point spread function was tested with our surface brightness fitting routine (see section \ref{sec:fitting}) and minor adjustments were made when necessary.

The typical full-width-half maximum of the stacked images was 0\farcs95, although it had to be adjusted from chip-to-chip. Chips further from the center of the mosaic generally had larger FWHM values than those near the center, but they never deviated by more than 10\% from this value. On a given chip, the variability of the point spread function was also modelled by DAOPHOT.  A `true' point spread function was derived for the entire chip, but several lookup tables were computed to measure deviations from this model. The point spread function was allowed to vary quadratrically over the chip and thus included terms proportional to $1, x, y, x^2, xy$ and $y^2$ where $x$ and $y$ are positions on the chip. The choice of the Moffat function to fit the stellar profile is fairly common. It is known to be numerically well behaved in fitting narrow point spread functions, and its wings also fit stellar profiles better than a gaussian profile \citep{moffat}.  The PSF fitting radius was 5\farcs6-6\farcs0. Larger values were found to contaminate the point spread function with neighbours, and smaller values did not provide a good model of the innermost pixels. 

\subsection{Bulge + Disk Decompositions}
\label{sec:fitting}
\begin{figure}[h!]
\plotone{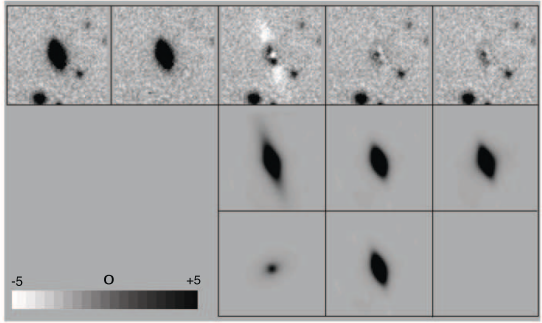}

\caption[An example of a bulge+disk fit]{Example of a bulge+disk fit. {\it Top row from left to right}: Postage stamp galaxy image extracted directly from science image, symmetrized galaxy image with neighbours removed, and residual images from the bulge, disk and bulge+disk model fits. {\it Middle row}: The $\chi^2$ values for the pure bulge, pure disk and bulge+disk models are 5.00, 1.38 and 1.21 respectively. {\it Bottom row}: Bulge and disk components of the bulge+disk model. In this case, as is true for the disks in our sample, the bulge component of the bulge+disk model is a small fraction of the total light for the galaxy. This galaxy has a bulge fraction of 9\%.} 

\label{fig:bd}
\end{figure}

\begin{figure}[h!]
\plotone{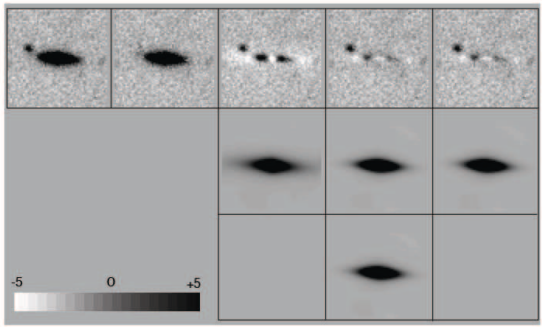}
\caption[An example of a disk fit]{Example of a pure disk fit. The layout of the images is same as Figure \ref{fig:bd}. This galaxy was best fit by a pure disk model.  As this image illustrates, the bulge+disk model ($\chi^2$=1.88) converged to a pure disk model ($\chi^2$=1.86). (Bulge model: $\chi^2$= 6.56) Although in this case the disk model was selected, both models yielded the same parameters for the galaxy, and the $\chi^2$ values between the models differed by less than 1\%. } 
\label{fig:disk}
\end{figure}

Measuring galaxy structural parameter from bulge+disk (B+D) image decompositions has been used extensively in the past \citep{simien,kent} and continues into the present day \citep{marleau,peng}. Here, we use the surface brightness fitting routine ``Galfit" described in detail in \citet{schade95} and \citet{saintonge}. This is not to be confused with another routine developed for structural decomposition of galaxy images also called 'GALFIT' and developed by \citet{galfit}.  For each of the 400,000 galaxies in our sample, ``Galfit" creates a 'postage stamp' of the object from the stacked image based on its \texttt{SExtractor} coordinates. Upon finding an accurate centroid for the object, a `symmetrized' image is generated in order to eliminate light contamination from neighbouring objects. Symmetrization is achieved by taking the original postage stamp, rotating it by 180$^\circ$, subtracting this rotated frame from the original and only preserving features that are at least 2$\sigma$ above the noise level of the image. Only these features that remain after the symmetrization process are used in the fitting process. The sky background level is not fit; an estimate of the background is input into the routine and then subtracted from the postage stamp. The symmetrized object used by Galfit is seen in Figures~\ref{fig:bd} and \ref{fig:disk} in the top row of images in the second panel from the left hand side. A detailed analysis of possible photometric biases due to symmetrization is given in \citet{saintonge}.

The routine convolves the point spread function with three idealized surface brightness models (pure $r^{1/4}$ bulge, pure disk and bulge+disk) and computes a $\chi^2$ for each. The routine fits 6 or 7 parameters for each of the idealized models. The actual set/number of fitting parameters depends on the particular model. For example, parameters for the bulge+disk model are bulge effective radius, bulge axial ratio, bulge position angle, disk scale length, disk axial ratio, disk position angle and fraction of bulge light to total galaxy light ($B/T$). 
The routine begins with preset initial parameter values and varies each parameter in the direction which decreases the $\chi^2$ of the fit. Initial parameters are globally set to some intermediate value (e.g., B/T always begins at 0.5, position angles at 45 degrees). The routine is sensitive to such input values only for objects that are below the detection limit of the survey; otherwise it generally converges to the same best-fit model regardless of initial values. The global minimum $\chi^2$ is determined by a modified Levenberg-Marquardt algorithm.

\section{Sample selection}\label{sampsel}

\subsection{Disk selection}\label{sec:disksel}

For every galaxy in the sample, Galfit returns $\chi^2$ values for the pure bulge, pure disk and bulge+disk model fits to the image. Objects that did not converge to a solution were not used in the sample described below. Approximately 4.1\% of the initial set of objects did not converge. Many of these were objects whose neighbours could not be properly removed or had other structural peculiarities. Some of these are discussed in \citet{saintonge}. We selected galaxies that had: 1) the lowest $\chi^2$  produced by a pure disk model or 2) the lowest $\chi^2$ produced by a bulge+disk model and $B/T \leq$ 0.2. The resulting sample includes nearly 65,000 galaxies plotted by redshift in Figure \ref{figs:sbsel}. Our disk selection must also include a lower cut on disk size to exclude galaxies with unreliable disk scale lengths that may affect the size function. In order to determine the smallest apparent galaxy size that could be reliably measured, we used a set of 5000 artificially generated galaxy images. 

The input objects were bulge+disk models with varying bulge fractions. The artificial galaxies were convolved with the real point spread function and then inserted into real image stacks. They were then analyzed with Galfit in exactly the same way as real galaxies with pure bulge, pure disk and bulge+disk models. We first compared the number of objects returned by Galfit versus the number of input objects as a function of the input disk scale length. We found that the number of simulated galaxies with recovered disk scale lengths less than $0\farcs37$ was as high as 10 times in excess of the input number. We carefully examined many ``problem" objects to visually confirm the effect. The effect was particularly problematic for sizes smaller than approximately $0\farcs21$. For sizes between $0\farcs21$ and $0\farcs32$, the fraction of objects returned improved drastically; Galfit would return a very reasonable 1.1-1.5 times the number of input objects at $0\farcs32$. However, at sizes greater than $0\farcs37$  the number returned matched the input number exactly. We therefore restricted our final disk sample to disks larger than this threshold to obtain a reliable size function. Our minimum threshold disk scale length corresponds to 3 kpc at $z$ = 1. We also used simulations to further characterize the region in size and luminosity over which disks could be reliably detected and measured out to $z$ = 1. Given our magnitude limit of $i'$=24.5, surface brightness dimming with increasing redshift pushes galaxies out of our sample, and larger disks are preferentially excluded because their light is spread out over a greater area. Our luminosity and size selection criteria are shown in Figure \ref{figs:sbsel}.

\subsection{Selection Effects}\label{sec:selectioneffects}

In order to evaluate the importance of selection effects on the calculation of the disk size function, the boundaries of the survey selection window must be carefully characterized. The treatment of selection effects can be done by adopting two different approaches. These approaches are not at all equivalent-  though they have been presented as such in previous studies of high-redshift disks. One approach is to determine the selection window of a survey at its high redshift end (where selection effects are expected to be most severe) and to apply this fixed selection window at all redshifts to map changes in galaxy properties within it. The other approach is to attempt to track the evolution of a given galaxy population across all redshifts by using a selection window that is varying with redshift based on a set of assumptions (e.g., pure luminosity evolution) regarding the expected evolution of the population of interest. An illustrative analogy to these two approaches is the two ways in which one can formulate the equations of fluid dynamics. The ``Eulerian" formulation uses coordinates that describe what happens at a fixed point in a fluid volume whereas the ``Lagrangian" formulation uses coordinates that follows a given fluid volume element as it moves around space. The ``Eulerian" \citep[e.g.,][]{simard99, ravi} and ``Lagrangian" \citep[e.g.,][]{schade96a, barden, sargent} treatments of selection effects in galaxy surveys have both been used in previous studies even  though they have been presented using the same language, and this ambiguity may be responsible for previous studies yielding results that appear discrepant at first sight.

Here we will use both approaches to the treatment of selection effects in our survey to illustrate how important the choice of approach is to the interpretation of the results from the same set of data. Unfortunately, neither approach is perfect. The ``Eulerian" approach may be tracking different populations as they move in and out of the fixed visibility window whereas the ``Lagrangian" approach may mistake trends in a non-evolving population for real evolution. This last point is particularly important. For example, local galaxy disks do not have a constant central surface brightness \citep[][Simard 2008, in prep.]{djl,shen2003,driver}, and the surface brightness of bright disks (the ones visible at $z=1)$ can be up to 1 magnitude brighter than the surface brightness of fainter disks. Such a variation in surface brightness could be interpreted as evolution by observing the same population at different redshifts with a selection window that does not look at the same range of disk luminosities.

\begin{figure}
\epsscale{1}
\plotone{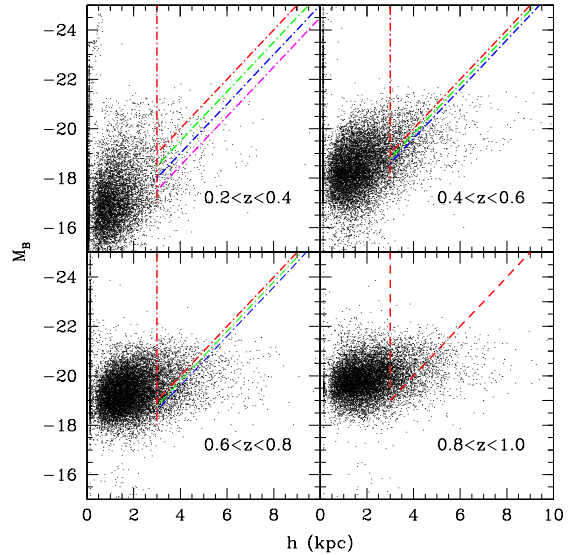}
\caption [Selected objects in luminosity size space]{Selected objects in disk luminosity-size space. {\it Dot-dashed red lines}:  selection window used to compute the size function for the no-evolution case. {\it Dashed coloured lines}: the `shifted' windows used to compute the size function with luminosity evolution. In the $0.6<z<0.8$ and $0.4<z<0.6$ bins, the green line indicates 0.2 mag of evolution, and the blue line indicates 0.4 mag. The $0.2<z<0.4$ bin shows the cases of 0.5, 1.0 and 1.5 mag of evolution shown in green, blue and violet, respectively.}
\label{figs:sbsel}
\end{figure}

To treat selection effects with a fixed visibility window, we begin by constraining the sample to the limits defined in the highest redshift bin. These limits are then applied to all lower $z$ bins as illustrated in Figure \ref{figs:sbsel}. Galaxies in the highest redshift bin ($0.8<z<1.0$) are the most constrained in luminosity-size space with respect to the other redshift bins as only the brightest galaxies remain detectable. Imposing the constraints in luminosity and size as a function of luminosity (surface brightness) present at high redshift on every redshift bin will ensure that the sample used here is free of surface brightness selection biases. This constraint ensures that comparable types of galaxies are used across all redshifts considered. The size function is computed only with the galaxies in this window.

Applying the same 'window' from the highest $z$ bin to all lower bins assumes no surface brightness evolution in the computation of the size function. In order to evaluate the effects of luminosity evolution on the size function, this window was moved by different amounts in magnitude as a function of redshift, and the size function was recomputed in each redshift bin. Different amounts of evolution were explored, and size functions were recomputed as a function of luminosity evolution. Various surface brightness models are plotted in Figure \ref{figs:sbsel}. They will be discussed later in section \ref{sec:sizesbev}.

\section{The Size Function}\label{the_sizefunc}
\label{sec:sf}
\indent The size function \citep{schade95,lilly98} is a powerful tool to characterize the size evolution of galaxies. The size function gives the space density of disks of a given scale  length at a given redshift.  It is important to emphasize that all previous works with the exception of  \citet{lilly98} and \citet{sargent} have focussed on the {\it zeropoint} of the luminosity-size relation of high-redshift disks. Therefore, they could not detect any evolution in luminosity and/or size that might have resulted in the same zeropoint. The size function provides a more stringent measure of disk evolution because it also places constraints on the {\it density} (i.e., distribution) of disks in luminosity and size.

Galaxies entering the calculation of the size function must first be weighted according to how well they are represented in the sample on the basis of their luminosity and redshift. Bright galaxies will be visible over larger cosmological volumes and would overwhelm galaxy samples if they were not also less abundant than faint ones. The appropriate weighted number densities are calculated using the 1/$V_{max}$ method as described in \citet{schmidt}. The accessible (or maximum) volume of a galaxy is defined as the comoving cosmological volume that a galaxy may occupy while remaining within the selection criteria of the survey. A galaxy of a given luminosity can be placed anywhere within this volume and still be detected. For the sake of simplicity, $V_{max}$ was computed with luminosity as the limiting factor as described below. Although the size of the galaxy may also determine the value of $z_{max}$\citep{shen2003}, we find here that luminosity is the limiting factor in greater than 99\% of cases and thus did not consider size limitations to $V_{max}$. To compute $V_{max}$, the sample is subdivided into a number of redshift shells. Let  $z_{min,s}$ and $z_{max,s}$ be the redshift limits of a shell, and let $z_{min,m}$ and $z_{max,m}$ be the redshift limits between which the galaxy meets the magnitude selection of the sample. Then the lower and upper redshift bounds for computing the maximum volume of a galaxy are $z_{min}$ = max($z_{min,s}$ , $z_{min,m}$) and $z_{max}$ = min($z_{max,s}$ , $z_{max,m}$).  One can integrate the comoving volume element between these redshift limits for each galaxy: 

\begin{eqnarray}
\label{eqn:vol}
V_{max} = \frac{D_H}{4\pi} \int_{\Omega_s} d\Omega f(\theta,\phi) \int_{z_{min}}^{z_{max}}\frac{D_C^2(z)}{\sqrt{\Omega_M (1+z)^3 + \Omega_{\Lambda}}} dz
\end{eqnarray}
where $D_H$ is the Hubble distance $c/H_0$, $H_0$ is the Hubble constant,  $\Omega_M$ is the parameter for matter density in the universe, and $\Omega_{\Lambda}$ is the cosmological constant \citep{hogg99}. The survey area $\Omega_s$ corresponding to our two CFHTLS-Deep fields is 1.83 square degrees.  The sampling fraction $f(\theta,\phi)$ as a function of position on the sky is constant over the survey area. $D_C(z)$ is the comoving distance defined as 
\begin{eqnarray}
\label{eqn:DC}
D_C(z) = D_H \int_{0}^{z} \frac{dz^{\prime}}{\sqrt{\Omega_M (1+z^{\prime})^3 + \Omega_{\Lambda}}}
\end{eqnarray}

Once 1/$V_{max}$ corrections are available for all galaxies in the sample, the space density of galaxies with disk scale lengths in the range $h$,$h+dh$ at redshift $z$ is then given by the sum:
\begin{eqnarray}
\label{eqn:sf}
\Phi (h,z) dh = \sum_{i = 1}^{N} \frac{1}{V_{max,i}}
\end{eqnarray}
\noindent where $h$ is the scale length of the disk, and $N$ is the number of disks in the sample at $z$. The size bin widths used here are 0.5 kiloparsecs. and size functions are computed in $\Delta z = 0.2 $ redshift shells ( $0.2 \leq z < 0.4$, $0.4 \leq z < 0.6$, $0.6 \leq z < 0.8$ and $0.8 \leq z < 1.0$). All error bars shown in the size functions presented here are computed using bootstrap resampling. The size function is computed 1000 times, each time using a different sample drawn from the original sample. The error bars represent 99\% confidence intervals based on the distribution of values from all the size function realizations, and they show the stability of the size function values within the data set.

\section{Results}\label{results}
\subsection{Size Function with Fixed Selection Window}
\label{sec:sizenoev}

\begin{figure}
\plotone{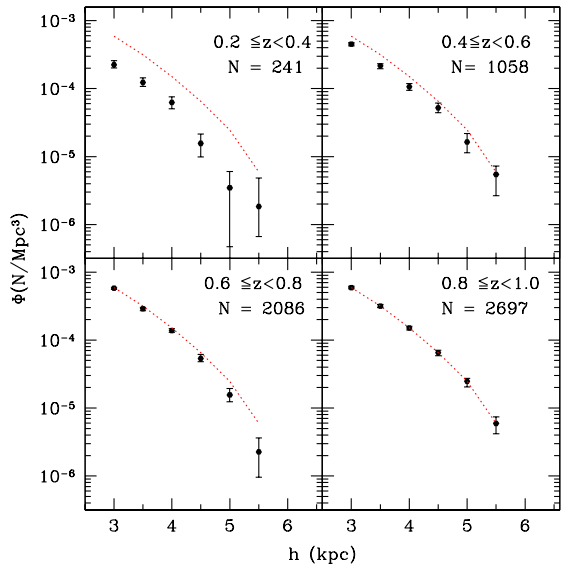}
\caption [Size function with no surface brightness evolution]{The size function for disks with $h>3$ kpc for the no evolution case. Galaxies included here reside within the $0.8<z<1.0$ selection window of Figure \ref{figs:sbsel}. The $0.8<z<1.0$ size function is shown as a red line in all the redshift bins for comparison.}
\label{figs:noev}
\end{figure}

The size function of disks as a function of redshift for the no-evolution case is shown in Figure~\ref{figs:noev}. It was computed by applying the same selection window at all redshift. The selection window is the one shown for the redshift range $0.8<z<1.0$ in Figure~\ref{figs:sbsel}. As mentioned earlier, this window places the most stringent limitations on size and luminosity of the galaxies used to calculate the size function. In this case, the size function is not tracking changes in a given disk population. Rather, it is measuring changes in the sizes of the disks that populate the fixed selection window at different redshifts.

\begin{figure}[h!]
\plotone{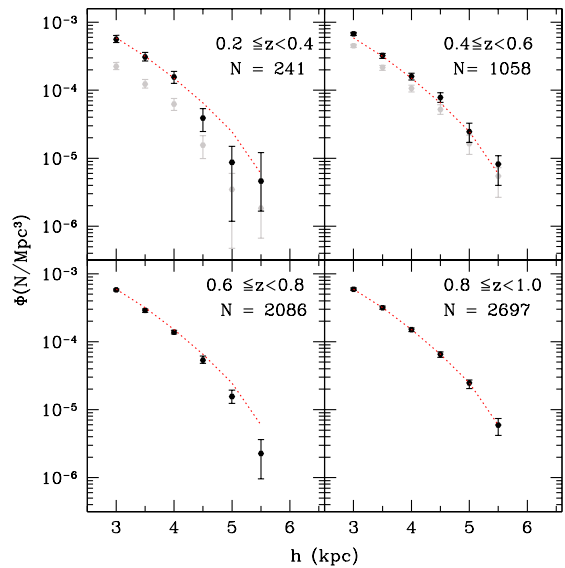}
\caption{Size function of disk galaxies with number density evolution in the lowest two redshift bins. The Size function at $0.2<z<0.4$ with a multiplicative factor of 2.5 applied is shown in black.  Size function at  $0.4<z<0.6$ with a factor of 1.5 applied is also shown in black.  The red line shows the size function at $0.8<z<1.0$ bin and is reproduced in all other redshift bins. Points in grey show the no-evolution size functions from Figure~\ref{figs:noev}.}
\label{figs:numdens}
\end{figure}

The shape of the size function in Figure \ref{figs:noev} is quite similar in all redshift slices. This suggests that the relative distribution of sizes is essentially constant over this redshift range i.e., there is no size-dependent size evolution.  The size functions of the two highest redshift bins seem to be in good agreement, but there is a significant difference between the size function in the lowest and highest redshift bins. This difference appears as a change in the normalization of the size function. Such a change may at first glance be indicative of a change in the number density of disks with redshift; however Figure~\ref{figs:numdens} shows that a pure number density evolution model would require about 2.5 times more disks at $0.8<z<1.0$ than at $0.2<z<0.4$ and 1.5 times more disk at $0.8<z<1.0$ than at $0.4<z<0.6$ in order to reproduce the observed size functions. The plausibility of such a model is discussed in Section~\ref{sec:purenum-evl}.

\subsection{Size Function with Varying Selection Window}\label{sec:sizesbev}

\begin{figure}[h!]
\plotone{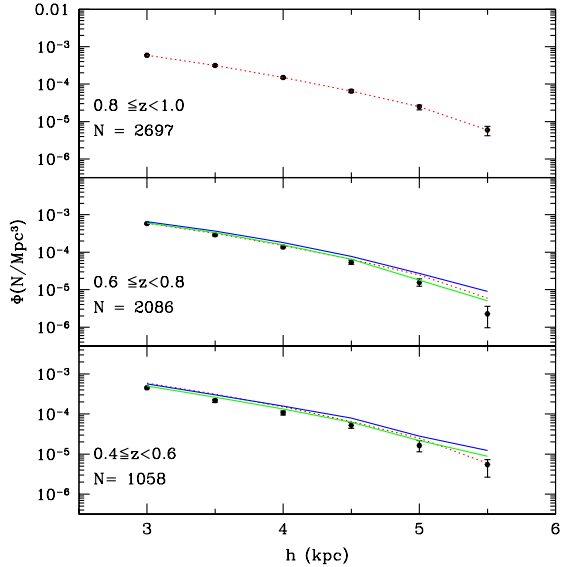}
\caption{Size function of disk galaxies with luminosity evolution. The red dashed line shows the size function at $0.8<z<1.0$ bin, and it is reproduced in all other redshift bins. The green and blue lines are the size functions at the indicated redshift with 0.2 and 0.4 mag of luminosity evolution respectively. Their selection windows are shown in Figure \ref{figs:sbsel}. Black points are the no-evolution size functions.}
\label{figs:evmodel}
\end{figure}

In order to compute the size function with a varying selection window, we first specify our constraints in size-magnitude space at $0.8<z<1.0$ as described earlier, and subsequently slide this window towards fainter absolute magnitudes with decreasing redshift until the size function at each redshift matches the high redshift one.  This sliding window is illustrated in Figure~\ref{figs:sbsel} by the coloured lines. This is equivalent to saying that disks are fading with time. Figure~\ref{figs:evmodel} shows size functions for the $0.4<z<0.6$ and $0.6<z<0.8$ bins with 0.2 and 0.4 mag of fading. These models do little to the size function at smaller sizes ($h<$ 4 kpc) but have some effect at larger sizes. About 0.2 mag is required to have the $0.6<z<0.8$ size function match the $0.8<z<1.0$ one, and 0.4 mag is needed  from $0.4<z<0.6$ to $0.8<z<1.0$. 

\begin{figure}[h!]
\plotone{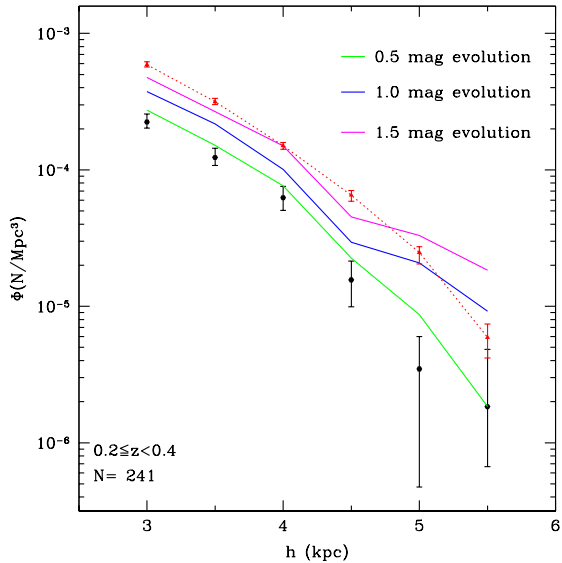}
\caption{Size function of low-z disk galaxies with pure luminosity evolution. The coloured lines are the $0.2<z<0.4$ size function with 0.5 (green), 1.0 (blue) and 1.5 (magenta) mag of evolution. Error bars for these models are the same as the no-evolution data points. The selection window for each size function is shown in Figure \ref{figs:sbsel} with the same corresponding colours. Black points are the the no-evolution $0.2<z<0.4$ size function, and the red triangles are the $0.8<z<1.0$ no-evolution size function.}
\label{figs:lowz}
\end{figure}

The $0.2<z<0.4$ size function is shown in Figure \ref{figs:lowz} with luminosity evolution ranging from 0.0 to 1.5 mag. Although neither the 0.5 mag nor no-evolution models can match this size function to the one at high redshift, about 1.0 and 1.5 mag appear to be adequate depending on disk size. The 1.5 mag model is a better model for sizes $\lesssim$ 4 kpc, and the 1.0 mag model is more reasonable at larger sizes. It is therefore possible that the largest galaxies may undergo a milder evolution than smaller galaxies. 

\subsection{Size Function from the Sloan Digital Sky Survey}\label{sec:sizef_sdss}

Studies of disks at high redshift can now greatly benefit from our increased knowledge of local disk properties thanks to the Sloan Digital Sky Survey (SDSS). A large dataset such as SDSS provides the ideal local baseline for comparison. We computed disk size functions from SDSS to compare with our CFHTLS results. The data come from Data Release Six \citep{adelman07} and the details of the SDSS size function analysis are given in Simard (2008, in preparation). Briefly, the three main galaxy selection criteria are Petrosian magnitude $r_{p,corr}$ (corrected for Galactic extinction), redshift and spectral classification. We selected objects with $ 14.0 \leq r_{p,corr} \leq 17.7$, $0.005 \leq z \leq 0.2$ and the spectrum of a galaxy as defined by the keyword {\tt SpecClass} in the {\tt SpecPhoto} database table ({\tt SpecClass}=2).  These criteria yielded 522,453 galaxies. The nominal $r-$band surface brightness limit of the SDSS spectroscopic sample is $\mu_{50}$ = 24.5 mag arcsec$^{-2}$ \citep{strauss}. However, we set the faint surface brightness limit of our sample to $\mu_{50}$ = 23.0 mag arcsec$^{-2}$ following \citet{shen2003} to retain a complete sample. The final total number of objects satisfying our selection criteria was 493,366. The  redshift distribution of this SDSS subsample peaks at $z \simeq $ 0.1.

\begin{figure*}
\includegraphics[angle=270,width=8.7cm]{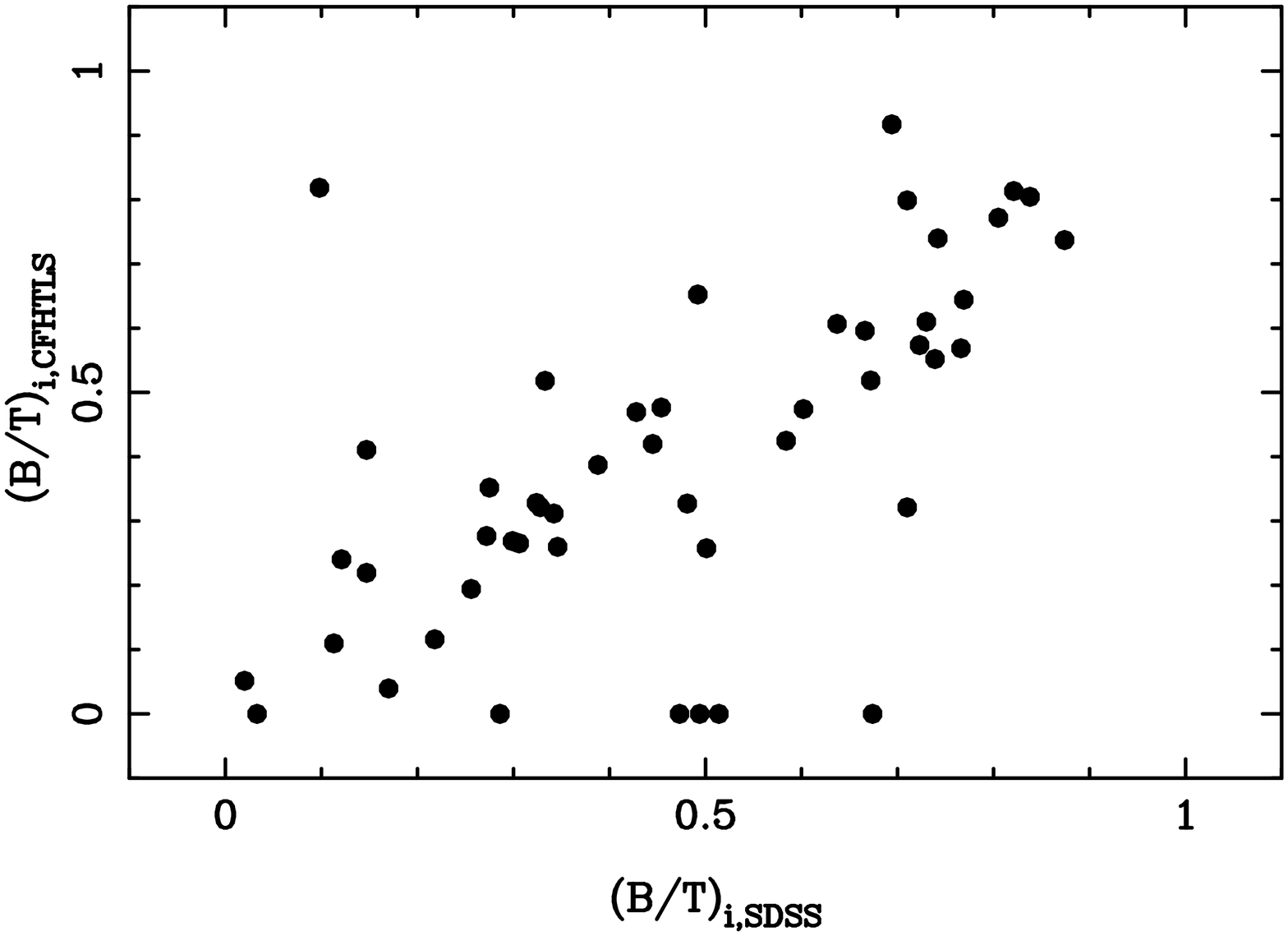}
\hfill
\includegraphics[angle=270,width=8.7cm]{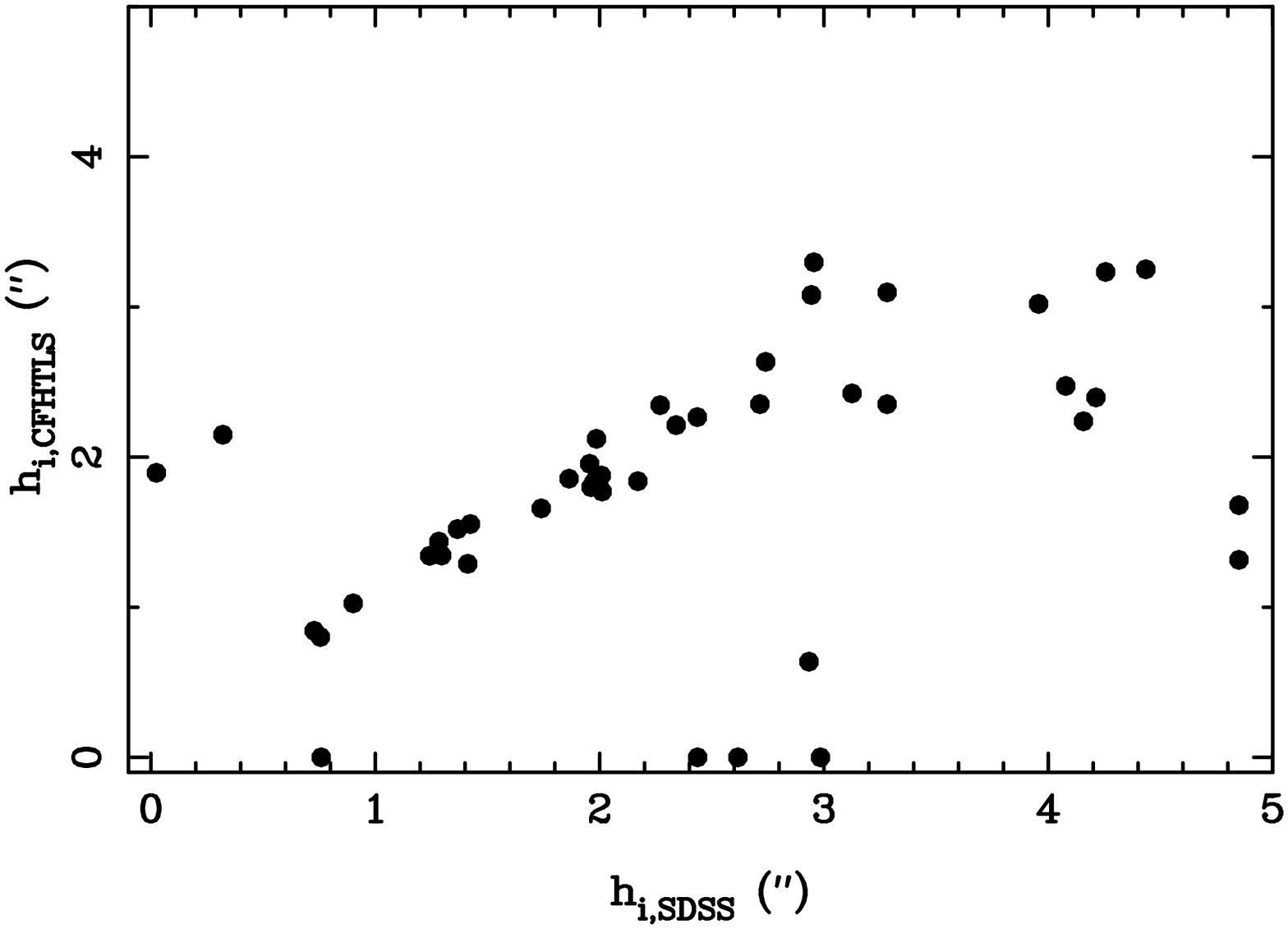}
\caption{Direct comparison between Galfit/CFHTLS and GIM2D/SDSS structural parameters. {\it Left-hand panel:} $i$-band bulge fraction. {\it Right-hand panel:} $i$-band disk scale length in arcseconds. All SDSS data are from Data Release 6.}
\label{figs:cfhtls_sdss_cmp}
\end{figure*}

Galaxy structural parameters were measured from bulge+disk decompositions performed using the GIM2D version 3.1 software package\citep{simard02}. We used the sum of an exponential disk and a de Vaucouleurs bulge (S\'ersic index $n = 4$) as our galaxy image model.  Fits in $g$, $r$ and $i$ were done using the separate fitting procedure described in \citet{simard02}.  GIM2D fitting failed for 226 objects (failure rate of 0.046$\%$),  but these ``objects" were artifacts (e.g., image defects, bright star spikes) in the SDSS catalog. Galfit was not run on the SDSS images, but we were able to directly compare GIM2D/SDSS and Galfit/CFHTLS bulge fractions and disk scale lengths using 49 galaxies in the overlap region between the CFHTLS D3 field and the SDSS survey area. The comparison was done in the $i$-band because the typical redshift of the SDSS galaxies is 0.1. The agreement is excellent as shown in Figure~\ref{figs:cfhtls_sdss_cmp}.    

\begin{figure*}
\includegraphics[angle=270,width=8.7cm]{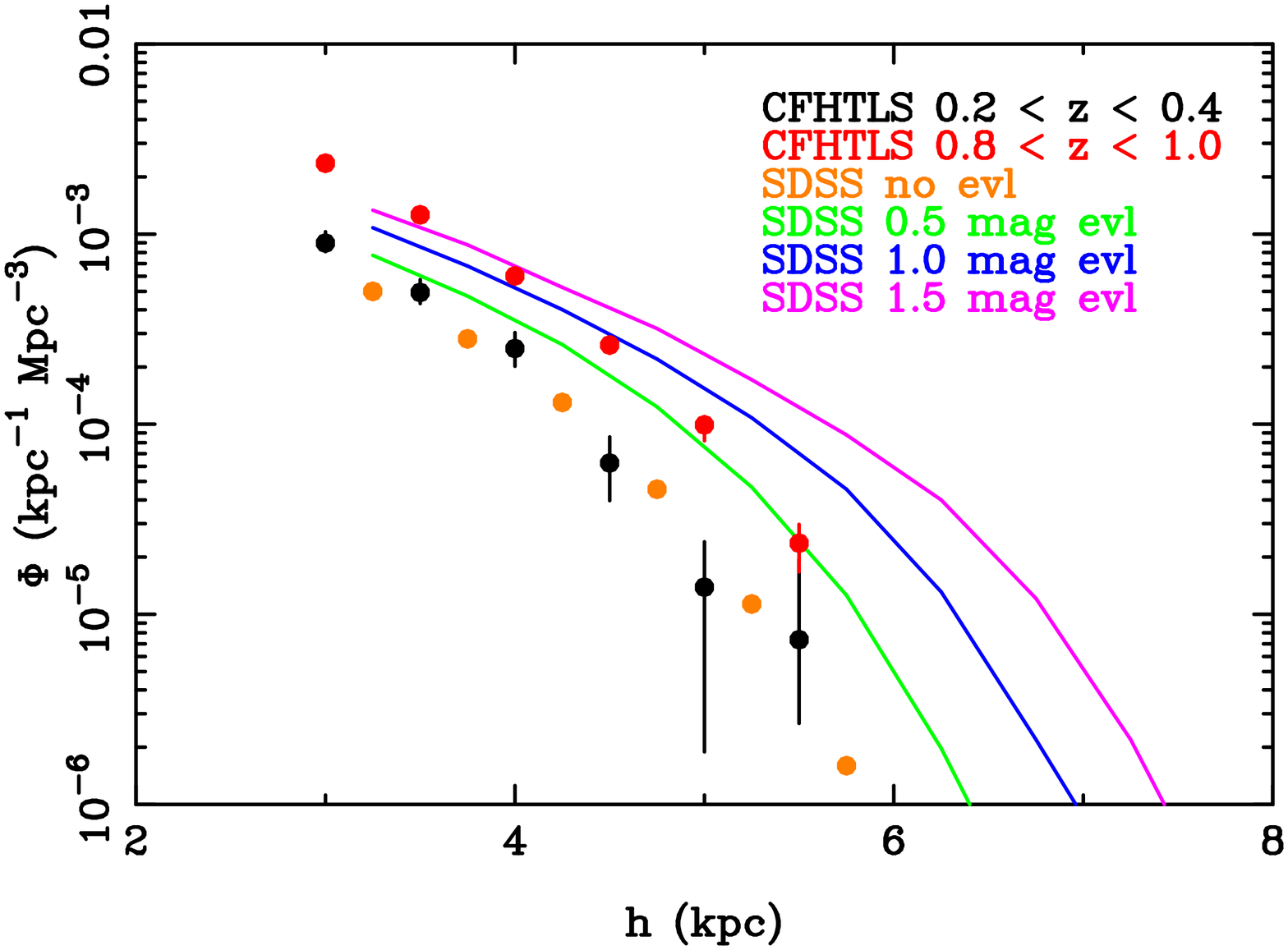}
\hfill
\includegraphics[angle=270,width=8.7cm]{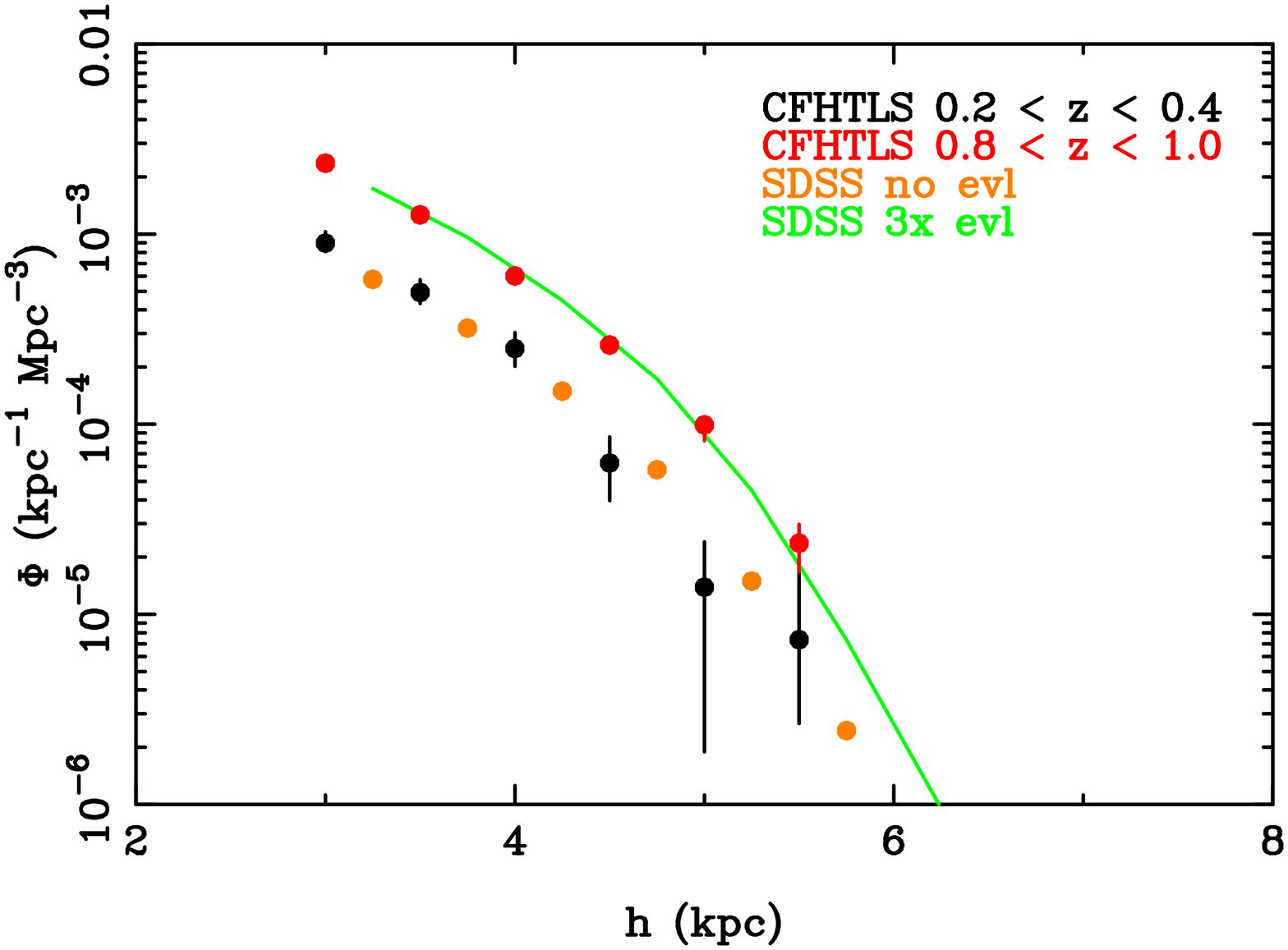}
\caption{Comparison between CFHTLS and SDSS disk size functions. 1$\sigma$ errors on the SDSS size functions in both figures are much smaller than the size of the data points. {\it Left-hand panel:} Pure luminosity evolution. The SDSS size functions with 0.0, 0.5, 1.0 and 1.5 magnitude of evolution include 41,056, 59,060, 74,747 and 86,832  disks respectively. {\it Right-hand panel:} Pure number density evolution. All SDSS data are from Data Release 6.}
\label{figs:sdss_sizefunc}
\end{figure*}

The comparison between the SDSS and CFHTLS size functions is shown in Figure~\ref{figs:sdss_sizefunc}. The SDSS size function is based on the $g$- band fits for all redshifts which corresponds to the $i$-band at $z = 0.9$. Given that size changes as a function of colour are small \citep[]{barden, color}, the effect of using a uniform filter for the respective CFHTLS and SDSS fits are not expected to have a significant effect on the size functions. The CFHTLS size functions are the no-evolution ones computed using the $0.8 < z < 1.0$ luminosity-size selection window from Figure~\ref{figs:sbsel}, and the no-evolution SDSS size function was computed using the same selection window. There is excellent agreement between the $0.2 < z < 0.4$ CFHTLS and SDSS ($z \simeq$ 0.1) size functions. This is a good check of the two independent size function calculations. The SDSS size functions with pure luminosity evolution were calculated by brightening the magnitudes of the SDSS galaxies before applying the CFHTLS selection window. This artificial brightening effectively selects the same objects in the magnitude-size plane as the varying selection window described earlier. Galaxies with $h \lesssim$ 4.5 kpc appear to require about 1.0$-$1.5 mag of brightening between $z = 0.1$ and $z = 0.9$ whereas 0.5 mag of brightening is a better match for the larger galaxies. This is generally the same trend seen in the CFHTLS data. The trend is more apparent in the SDSS size functions because they have much better statistics. The SDSS size function with pure number density evolution was calculated by simply multiplying the SDSS space densities by a factor of 3 before applying the CFHTLS selection window. It provides a very good match to the $0.8 < z < 1.0$ CFHTLS size function.  

\section{Discussion}\label{discussion}

The fundamental variables of disk evolution are luminosity, size and number density.  Some studies have measured changes in the zeropoint of the luminosity-size relation (i.e., surface brightness), and it is important to emphasize that they could not disentangle size and luminosity evolution. Other studies provided stronger constraints by including  number density information through the size function. The list of previous studies on disk evolution is given in Table~\ref{comptable}. We divide the following discussion into pure luminosity and pure number density evolution for the sake of clarity with the understanding that in reality, both variables affect the population to some degree.

\begin{table*}
\caption[]{Comparison of surface brightness evolution of different works}
\begin{center}
\begin{tabular}{lccc}
\hline
Work & Number of disks & Redshift Range & $\mu$ evolution \\
\hline
Forbes et al., 1996 & 17 & $0.20 \le z \le 0.84$ & $0.6-0.85$ \\
Roche et al., 1998 & 270\tablenotemark{a} & $0.20 \le z \le 0.90$   & 0.94 $\pm$ 0.23 \\
Lilly et al., 1998 & 341 & $0.10 \le z \le 1.00$  & 0.8 $\pm$ 0.3 \\
Schade et al., 1996a & 110 & $0.50 \le z \le 1.10$  & 1.6 $\pm$ 0.1 \\
Simard et al., 1999 & 136 & $0.00 \le z \le 1.00 $  & minimal  \\
Ravindranath et al., 2004 & 1508 & $0.20 \le z \le 1.25 $  & $\lesssim$ 0.4  \\
Barden et al., 2005 & 5506 & $0.00 \le z \le 1.10 $ & 0.99 $\pm$ 0.06\tablenotemark{b} \\
Sargent et al., 2007 & 12000 & $0.00 < z < 1.00$ & 1.0 \\
This Work & 65000 & $0.20 \le z \le 1.00 $ & $1.0-1.5$ \\
\hline
\end{tabular}

\tablenotetext{a}{to $z$ = 3.43}
\tablenotetext{b}{V band}
\end{center}
\label{comptable}
\end{table*}

\subsection{Pure Luminosity Evolution}\label{sec:purelum-evl}

The amount of luminosity evolution found by all previous works ranges from 0 to 1.6 mag. \citet{simard99} and \citet{ravi} found little or no surface brightness evolution and pointed out that an artificial evolutionary effect could be introduced by selecting galaxies in different ranges of absolute magnitudes at different redshifts. This artificial effect was due to local galaxies not following a constant brightness relation \citep[][Simard 2008, in prep.]{djl,shen2003,driver}. Local disks with absolute magnitudes bright enough to be visible at high redshift are nearly 1 mag brighter in surface brightness than fainter disks. In section \ref{sec:sizenoev}, we selected objects in the same manner as these groups and found that number density evolution was needed in order to reproduce observed size functions. We will revisit this in the next section. 

\citet{lilly98} and \citet{sargent} found that the size function of disks was constant with redshift after 1 mag of luminosity evolution had been included. This is in agreement with what we found for the case of a varying selection window. Our own estimates are between 1.0 and 1.5 mag with a weak dependence on size. As seen in Figure \ref{figs:lowz}, the 1.5 mag evolution model seems to be too much for the largest disks in our sample ($h > 4.5$ kpc). The 1 mag model provides a better match for these larger disks. Our comparison between SDSS and CFHTLS size functions yielded a somewhat lower amount of evolution: 1 mag for disks with $h \lesssim$ 4.5 kpc and 0.5 mag for the larger ones. 

The luminosity evolution of a disk is obviously tied to the evolution of its constituent stellar populations. It is therefore interesting that even the largest amount of luminosity evolution is lower than expected from passive evolution of a stellar population. Passive evolution is thought to be responsible for at least 1.5 mag of evolution since $z=1.3$ in early type galaxies \citep{vandokkum}, and perhaps more \citep{treu, gebhardt}. An estimate of 0.7 mag between $0.0 < z < 0.7$ from \citet{cassata} is in rough agreement with our sample. However, the disks examined here may have had a more active star formation in the past than the galaxies of these studies. This would suggest that our evolution estimate is likely lower than expected, although this amount is highly dependent on star formation history and the growth of stellar mass. The star formation history of the universe has declined sharply since $z = 1$ \citep{madau}, and star formation rate is highly dependent on mass \citep{juneau05}: less massive (and smaller) systems have been far more active in recent times than more massive ones. The pure luminosity evolution model here assumes that {\it all} disks have undergone the same amount of evolution. A greater luminosity evolution for smaller galaxies as seen in our data and previous works could be interpreted in terms of increased activity in lower mass systems \citep[e.g.,][]{zheng}.

\subsection {Pure Number Density Evolution}\label{sec:purenum-evl}

In section \ref{sec:sizenoev}, we presented the size function of galaxies computed with a fixed selection window at all redshifts. The size functions at $0.2<z<0.4$ and  $0.8<z<1.0$ could be reconciled by increasing the number density of high redshift disks by a factor of 2.5 with no size evolution (Figure~\ref{figs:numdens}) . A comparison with SDSS required an increase by a factor of 3 (Figure~\ref{figs:sdss_sizefunc}). \citet{sargent} found that larger disks were less abundant at high redshift albeit with a more modest increase in density;  the largest disks in their sample ($h>$10 kpc) were only 60\% as abundant as local disks. In determining what amount of number density evolution is actually needed for our sample, it is important to remember that we selected the disks of the most disk-dominated galaxies ($B/T \leq$ 0.2), and that disks that have recently grown a significant a bulge component would be excluded.  The growth of `classical' bulges is driven by mergers, whereas the growth of `pseudobulges' is driven by internal (secular) processes within the galaxy. 

Mergers and the quenching of star formation are responsible for transferring galaxies from the so-called ``blue cloud" where most of the disk-dominated galaxies in our sample would reside to the ``red sequence" of quiescent galaxies \citep{bell2006,faber}. Although disks are relatively fragile, they are not entirely destroyed in merger events. Disks may re-form in merger remnants if the merging galaxies were sufficiently gas-rich  \citep{springer05,robertson,naab06,brook07}. However, all merger remnants would all have bulge components that would force them out of our sample.  The epoch of merger-driven galaxy evolution is thought to have peaked early ($z \sim 2-3$) in the history of the Universe \citep{kormkenn}. The merger rate in recent times, particularly since $z=1$, is thought to have been quite low  \citep{patton,bundy,lin,blanton}. More specifically, \citet{patton} find that only 6.6\% of present day galaxies with rest-frame $B$-band absolute magnitudes $-18< M_B <-21$ (similar to the range probed by our sample) have had a major merger since $z=1$. Galaxies in the blue cloud are thought to follow a relatively quiet evolution with less than 10\% experiencing a major merger which would then place them on the red sequence \citep{blanton}. Therefore, such a small fraction of mergers cannot yield significant number density evolution. Although the merger rates for more massive galaxies ($M_* > 5 \times 10^{10} M_\odot$) may be slightly higher \citep{bell2006}, it is unclear how this would relate to our disk sample given that it applies to a wide range of morphologies and mass-to-light ratios.  

As noted above, mergers are not be the only mechanism that can transform pure disk galaxies into more bulge dominated ones. As described in \citet{Bower}, field galaxies can undergo secular evolution processes which may produce a significant change in the number density of disks since $z=1$.  Disk instabilities and bars are examples of dynamical drivers which would cause significant morphological transformation without external influence. In these cases, gas within the disk is redistributed towards the centre leading to enhanced star formation and bulge growth \citep{kormkenn}. The resulting ``pseudobulges"  have dynamically cold structures similar to disks, and their host galaxies are typically located within the blue cloud \citep{drory}.  The timescale for the formation of a bar and then a pseudobulge is a few gigayears \citep{kormkenn,drory}, and the time elapsed since $z = 1$ would be amply sufficient to allow this process to operate on the disk population.

\section{Conclusions}\label{sec:conclusions}

Using deep, wide-field imaging from the CFHT Legacy Survey covering two square degrees, we perform two-dimensional, bulge+disk decompositions on all galaxies brighter than $i^{\prime} \leq$ 24.5. Biases are carefully characterized through artificial galaxy simulations. We select the disks of the most disk-dominated galaxies with bulge fraction less than 20$\%$ (6000 galaxies), and we compute disk size functions over the redshift range $0.2 < z <1.0$ using different selection windows to illustrate the importance of selection effects. We also compute the size function of a very large sample of disks from the Sloan Digital Sky Survey to use as a local  ($z \simeq$ 0.1) comparison. We find that:

1. The CFHTLS size functions computed by using the same fixed luminosity-size selection window at all redshifts exhibit evolution that appears to be best modelled by a pure number density evolution. The $z = 0.3$ size function is an excellent match to the $z = 0.9$ one if disks at the highest redshift are a factor of 2.5 more abundant than in the local universe. The SDSS size function would also match the $z = 0.9$ CFHTLS size function very well with a similar change in number density. However, current estimates of the galaxy merger rates below $z = 1$ are too low to produce such a significant change in the number density of the disk population in our sample.

2. The CFHTLS size functions computed by using a varying luminosity-size selection window with redshift remain constant if the selection window is shifted by 1.0$-$1.5 mag towards fainter magnitudes with decreasing redshift. There is a weak dependence on disk scale length with smaller ($h \lesssim 4$ kpc) disks requiring more luminosity evolution than larger ones. This amount of pure luminosity evolution is consistent with previous works and expectations from the evolution of disk constituent stellar populations.

The evolution of disks is likely to be a more complicated interplay between size, luminosity and number density than the simple pure luminosity and pure number density evolution models explored here. Models of disk formation and growth are still in their infancy. Observational results are beginning to provide interesting constraints, but more work remains to be done. Direct measurements of disk spectral energy distributions through full multi-color bulge+disk decompositions over a wide range of wavelengths would go a long way in furthering this effort as it would disentangle the evolution of disks as a function of morphology, mass-to-light ratio and  stellar mass.

\acknowledgments

LS gratefully acknowledges the receipt of a Discovery Grant from the Natural Science and Engineering Research
Council of Canada which funded some of this research. This research made use of a University of Victoria computing facility funded by grants from the Canadian Foundation for Innovation and the British Columbia Knowledge and Development Fund.  Funding for the SDSS and SDSS-II has been provided by the Alfred P. Sloan Foundation, the Participating Institutions, the National Science Foundation, the U.S. Department of Energy, the National Aeronautics and Space Administration, the Japanese Monbukagakusho, the Max Planck Society, and the Higher Education Funding Council for England. The SDSS Web Site is http://www.sdss.org/. The SDSS is managed by the Astrophysical Research Consortium for the Participating Institutions. The Participating Institutions are the American Museum of Natural History, Astrophysical Institute Potsdam, University of Basel, University of Cambridge, Case Western Reserve University, University of Chicago, Drexel University, Fermilab, the Institute for Advanced Study, the Japan Participation Group, Johns Hopkins University, the Joint Institute for Nuclear Astrophysics, the Kavli Institute for Particle Astrophysics and Cosmology, the Korean Scientist Group, the Chinese Academy of Sciences (LAMOST), Los Alamos National Laboratory, the Max-Planck-Institute for Astronomy (MPIA), the Max-Planck-Institute for Astrophysics (MPA), New Mexico State University, Ohio State University, University of Pittsburgh, University of Portsmouth, Princeton University, the United States Naval Observatory, and the University of Washington.

\bibliography{ms}

\begin{thebibliography}{}

\bibitem[\protect\astroncite{{Adelman-McCarthy et al.}}{2007}]{adelman07}
{Adelman-McCarthy et al.}, J.~K.: 2007,
\newblock {\em ArXiv e-prints} 707

\bibitem[\protect\astroncite{{Barden} et~al.}{2005}]{barden}
{Barden}, M., {Rix}, H.-W., {Somerville}, R.~S., {Bell}, E.~F.,
  {H{\"a}u{\ss}ler}, B., {Peng}, C.~Y., {Borch}, A., {Beckwith}, S.~V.~W.,
  {Caldwell}, J.~A.~R., {Heymans}, C., {Jahnke}, K., {Jogee}, S., {McIntosh},
  D.~H., {Meisenheimer}, K., {S{\'a}nchez}, S.~F., {Wisotzki}, L., and {Wolf},
  C.: 2005,
\newblock {\em ApJ} {\bf 635}, 959

\bibitem[\protect\astroncite{{Barnes} and {Hernquist}}{1992}]{barnes}
{Barnes}, J.~E. and {Hernquist}, L.: 1992,
\newblock {\em \araa} {\bf 30}, 705

\bibitem[\protect\astroncite{{Bell} et~al.}{2006}]{bell2006}
{Bell}, E.~F., {Phleps}, S., {Somerville}, R.~S., {Wolf}, C., {Borch}, A., and
  {Meisenheimer}, K.: 2006,
\newblock {\em ApJ} {\bf 652}, 270

\bibitem[\protect\astroncite{{Bertin} and {Arnouts}}{1996}]{bertin}
{Bertin}, E. and {Arnouts}, S.: 1996,
\newblock {\em A\&APS} {\bf 117}, 393

\bibitem[\protect\astroncite{{Blanton}}{2006}]{blanton}
{Blanton}, M.~R.: 2006,
\newblock {\em \apj} {\bf 648}, 268

\bibitem[\protect\astroncite{{Bower} et~al.}{2006}]{Bower}
{Bower}, R.~G., {Benson}, A.~J., {Malbon}, R., {Helly}, J.~C., {Frenk}, C.~S.,
  {Baugh}, C.~M., {Cole}, S., and {Lacey}, C.~G.: 2006,
\newblock {\em \mnras} {\bf 370}, 645

\bibitem[\protect\astroncite{{Brook} et~al.}{2007}]{brook07}
{Brook}, C., {Richard}, S., {Kawata}, D., {Martel}, H., and {Gibson}, B.~K.:
  2007,
\newblock {\em apj} {\bf 658}, 60

\bibitem[\protect\astroncite{{Bundy} et~al.}{2004}]{bundy}
{Bundy}, K., {Fukugita}, M., {Ellis}, R.~S., {Kodama}, T., and {Conselice},
  C.~J.: 2004,
\newblock {\em ApJL} {\bf 601}, L123

\bibitem[\protect\astroncite{{Cassata} et~al.}{2007}]{cassata}
{Cassata}, P., {Guzzo}, L., {Franceschini}, A., {Scoville}, N., {Capak}, P.,
  {Ellis}, R.~S., {Koekemoer}, A., {McCracken}, H.~J., {Mobasher}, B.,
  {Renzini}, A., {Ricciardelli}, E., {Scodeggio}, M., {Taniguchi}, Y., and
  {Thompson}, D.: 2007,
\newblock {\em \apjs} {\bf 172}, 270

\bibitem[\protect\astroncite{{Coleman} et~al.}{1980}]{cww}
{Coleman}, G.~D., {Wu}, C.-C., and {Weedman}, D.~W.: 1980,
\newblock {\em \apjs} {\bf 43}, 393

\bibitem[\protect\astroncite{{de Jong}}{1996}]{color}
{de Jong}, R.~S.: 1996,
\newblock {\em \aap} {\bf 313}, 377

\bibitem[\protect\astroncite{{de Jong} and {Lacey}}{2000}]{djl}
{de Jong}, R.~S. and {Lacey}, C.: 2000,
\newblock {\em \apj} {\bf 545}, 781

\bibitem[\protect\astroncite{{Driver} et~al.}{2005}]{driver}
{Driver}, S.~P., {Liske}, J., {Cross}, N.~J.~G., {De Propris}, R., and {Allen},
  P.~D.: 2005,
\newblock {\em \mnras} {\bf 360}, 81

\bibitem[\protect\astroncite{{Drory} and {Fisher}}{2007}]{drory}
{Drory}, N. and {Fisher}, D.~B.: 2007,
\newblock {\em \apj} {\bf 664}, 640

\bibitem[\protect\astroncite{{Faber} et~al.}{2007}]{faber}
{Faber}, S.~M., {Willmer}, C.~N.~A., {Wolf}, C., {Koo}, D.~C., {Weiner}, B.~J.,
  {Newman}, J.~A., {Im}, M., {Coil}, A.~L., {Conroy}, C., {Cooper}, M.~C.,
  {Davis}, M., {Finkbeiner}, D.~P., {Gerke}, B.~F., {Gebhardt}, K., {Groth},
  E.~J., {Guhathakurta}, P., {Harker}, J., {Kaiser}, N., {Kassin}, S.,
  {Kleinheinrich}, M., {Konidaris}, N.~P., {Kron}, R.~G., {Lin}, L., {Luppino},
  G., {Madgwick}, D.~S., {Meisenheimer}, K., {Noeske}, K.~G., {Phillips},
  A.~C., {Sarajedini}, V.~L., {Schiavon}, R.~P., {Simard}, L., {Szalay}, A.~S.,
  {Vogt}, N.~P., and {Yan}, R.: 2007,
\newblock {\em \apj} {\bf 665}, 265

\bibitem[\protect\astroncite{{Fall} and {Efstathiou}}{1980}]{fall}
{Fall}, S.~M. and {Efstathiou}, G.: 1980,
\newblock {\em MNRAS} {\bf 193}, 189

\bibitem[\protect\astroncite{{Gebhardt} et~al.}{2003}]{gebhardt}
{Gebhardt}, K., {Faber}, S.~M., {Koo}, D.~C., {Im}, M., {Simard}, L.,
  {Illingworth}, G.~D., {Phillips}, A.~C., {Sarajedini}, V.~L., {Vogt}, N.~P.,
  {Weiner}, B., and {Willmer}, C.~N.~A.: 2003,
\newblock {\em \apj} {\bf 597}, 239

\bibitem[\protect\astroncite{{Gwyn}}{2008}]{gwyn}
{Gwyn}, S.~D.~J.: 2008,
\newblock {\em PASP, in press}

\bibitem[\protect\astroncite{{Hogg}}{1999}]{hogg99}
{Hogg}, D.~W.: 1999,
\newblock {\em ArXiv Astrophysics e-prints}

\bibitem[\protect\astroncite{{Juneau et al.}}{2005}]{juneau05}
{Juneau et al.}, S.: 2005,
\newblock {\em \apjl} 619

\bibitem[\protect\astroncite{{Kent}}{1985}]{kent}
{Kent}, S.~M.: 1985,
\newblock {\em \apjs} {\bf 59}, 115

\bibitem[\protect\astroncite{{Kormendy} and {Kennicutt}}{2004}]{kormkenn}
{Kormendy}, J. and {Kennicutt}, Jr., R.~C.: 2004,
\newblock {\em \araa} {\bf 42}, 603

\bibitem[\protect\astroncite{{Lilly} et~al.}{1998}]{lilly98}
{Lilly}, S., {Schade}, D., {Ellis}, R., {Le Fevre}, O., {Brinchmann}, J.,
  {Tresse}, L., {Abraham}, R., {Hammer}, F., {Crampton}, D., {Colless}, M.,
  {Glazebrook}, K., {Mallen-Ornelas}, G., and {Broadhurst}, T.: 1998,
\newblock {\em ApJ} {\bf 500}, 75

\bibitem[\protect\astroncite{{Lin} et~al.}{2004}]{lin}
{Lin}, L., {Koo}, D.~C., {Willmer}, C.~N.~A., {Patton}, D.~R., {Conselice},
  C.~J., {Yan}, R., {Coil}, A.~L., {Cooper}, M.~C., {Davis}, M., {Faber},
  S.~M., {Gerke}, B.~F., {Guhathakurta}, P., and {Newman}, J.~A.: 2004,
\newblock {\em ApJL} {\bf 617}, L9

\bibitem[\protect\astroncite{{Madau} et~al.}{1996}]{madau}
{Madau}, P., {Ferguson}, H.~C., {Dickinson}, M.~E., {Giavalisco}, M.,
  {Steidel}, C.~C., and {Fruchter}, A.: 1996,
\newblock {\em \mnras} {\bf 283}, 1388

\bibitem[\protect\astroncite{{Marleau} and {Simard}}{1998}]{marleau}
{Marleau}, F.~R. and {Simard}, L.: 1998,
\newblock {\em \apj} {\bf 507}, 585

\bibitem[\protect\astroncite{{Mo} et~al.}{1998}]{momao}
{Mo}, H.~J., {Mao}, S., and {White}, S.~D.~M.: 1998,
\newblock {\em MNRAS} {\bf 295}, 319

\bibitem[\protect\astroncite{{Naab} et~al.}{2006}]{naab06}
{Naab}, T., {Jessit}, R., and {Burket}, A.: 2006,
\newblock {\em MNRAS} {\bf 372}, 839

\bibitem[\protect\astroncite{{Nuijten} et~al.}{2005}]{nuij}
{Nuijten}, M.~J.~H.~M., {Simard}, L., {Gwyn}, S., and {R{\"o}ttgering},
  H.~J.~A.: 2005,
\newblock {\em \apjl} {\bf 626}, L77

\bibitem[\protect\astroncite{{Patton} et~al.}{2000}]{patton}
{Patton}, D.~R., {Carlberg}, R.~G., {Marzke}, R.~O., {Pritchet}, C.~J., {da
  Costa}, L.~N., and {Pellegrini}, P.~S.: 2000,
\newblock {\em ApJ} {\bf 536}, 153

\bibitem[\protect\astroncite{{Peng} et~al.}{2002a}]{peng}
{Peng}, C.~Y., {Ho}, L.~C., {Impey}, C.~D., and {Rix}, H.-W.: 2002a,
\newblock {\em \aj} {\bf 124}, 266

\bibitem[\protect\astroncite{{Peng} et~al.}{2002b}]{galfit}
{Peng}, C.~Y., {Ho}, L.~C., {Impey}, C.~D., and {Rix}, H.-W.: 2002b,
\newblock {\em \aj} {\bf 124}, 266

\bibitem[\protect\astroncite{{Ravindranath} et~al.}{2004}]{ravi}
{Ravindranath}, S., {Ferguson}, H.~C., {Conselice}, C., {Giavalisco}, M.,
  {Dickinson}, M., {Chatzichristou}, E., {de Mello}, D., {Fall}, S.~M.,
  {Gardner}, J.~P., {Grogin}, N.~A., {Hornschemeier}, A., {Jogee}, S.,
  {Koekemoer}, A., {Kretchmer}, C., {Livio}, M., {Mobasher}, B., and
  {Somerville}, R.: 2004,
\newblock {\em ApJL} {\bf 604}, L9

\bibitem[\protect\astroncite{{Rhodes} et~al.}{2000}]{rhodes2000}
{Rhodes}, J., {Refregier}, A., and {Groth}, E.~J.: 2000,
\newblock {\em \apj} {\bf 536}, 79

\bibitem[\protect\astroncite{{Robertson} et~al.}{2006}]{robertson}
{Robertson}, B., {Bullock}, J.~S., {Cox}, T.~J., {Di Matteo}, T., {Hernquist},
  L., {Springel}, V., and {Yoshida}, N.: 2006,
\newblock {\em \apj} {\bf 645}, 986

\bibitem[\protect\astroncite{{Roche} et~al.}{1998}]{roche98}
{Roche}, N., {Ratnatunga}, K., {Griffiths}, R.~E., {Im}, M., and {Naim}, A.:
  1998,
\newblock {\em MNRAS} {\bf 293}, 157

\bibitem[\protect\astroncite{{Saintonge} et~al.}{2005}]{saintonge}
{Saintonge}, A., {Schade}, D., {Ellingson}, E., {Yee}, H.~K.~C., and
  {Carlberg}, R.~G.: 2005,
\newblock {\em \apjs} {\bf 157}, 228

\bibitem[\protect\astroncite{{Sargent} et~al.}{2007}]{sargent}
{Sargent}, M.~T., {Carollo}, C.~M., {Lilly}, S.~J., {Scarlata}, C., {Feldmann},
  R., {Kampczyk}, P., {Koekemoer}, A.~M., {Scoville}, N., {Kneib}, J.-P.,
  {Leauthaud}, A., {Massey}, R., {Rhodes}, J., {Tasca}, L.~A.~M., {Capak}, P.,
  {McCracken}, H.~J., {Porciani}, C., {Renzini}, A., {Taniguchi}, Y.,
  {Thompson}, D.~J., and {Sheth}, K.: 2007,
\newblock {\em \apjs} {\bf 172}, 434

\bibitem[\protect\astroncite{{Schade} et~al.}{1996}]{schade96a}
{Schade}, D., {Carlberg}, R.~G., {Yee}, H.~K.~C., {Lopez-Cruz}, O., and
  {Ellingson}, E.: 1996,
\newblock {\em \apjl} {\bf 465}, L103+

\bibitem[\protect\astroncite{{Schade} et~al.}{1995}]{schade95}
{Schade}, D., {Lilly}, S.~J., {Crampton}, D., {Hammer}, F., {Le Fevre}, O., and
  {Tresse}, L.: 1995,
\newblock {\em ApJL} {\bf 451}, L1+

\bibitem[\protect\astroncite{{Schmidt}}{1968}]{schmidt}
{Schmidt}, M.: 1968,
\newblock {\em \apj} {\bf 151}, 393

\bibitem[\protect\astroncite{{Shen} et~al.}{2003}]{shen2003}
{Shen}, S., {Mo}, H.~J., {White}, S.~D.~M., {Blanton}, M.~R., {Kauffmann}, G.,
  {Voges}, W., {Brinkmann}, J., and {Csabai}, I.: 2003,
\newblock {\em \mnras} {\bf 343}, 978

\bibitem[\protect\astroncite{{Simard} et~al.}{1999}]{simard99}
{Simard}, L., {Koo}, D.~C., {Faber}, S.~M., {Sarajedini}, V.~L., {Vogt}, N.~P.,
  {Phillips}, A.~C., {Gebhardt}, K., {Illingworth}, G.~D., and {Wu}, K.~L.:
  1999,
\newblock {\em ApJ} {\bf 519}, 563

\bibitem[\protect\astroncite{{Simard} et~al.}{2002}]{simard02}
{Simard}, L., {Willmer}, C.~N.~A., {Vogt}, N.~P., {Sarajedini}, V.~L.,
  {Phillips}, A.~C., {Weiner}, B.~J., {Koo}, D.~C., {Im}, M., {Illingworth},
  G.~D., and {Faber}, S.~M.: 2002,
\newblock {\em \apjs} {\bf 142}, 1

\bibitem[\protect\astroncite{{Simien} and {de Vaucouleurs}}{1986}]{simien}
{Simien}, F. and {de Vaucouleurs}, G.: 1986,
\newblock {\em ApJ} {\bf 302}, 564

\bibitem[\protect\astroncite{{Somerville} et~al.}{2004}]{somerville04}
{Somerville}, R.~S., {Lee}, K., {Ferguson}, H.~C., {Gardner}, J.~P.,
  {Moustakas}, L.~A., and {Giavalisco}, M.: 2004,
\newblock {\em \apjl} {\bf 600}, L171

\bibitem[\protect\astroncite{{Springer} and {Hernquist}}{2005}]{springer05}
{Springer}, V. and {Hernquist}, L.: 2005,
\newblock {\em ApJ} {\bf 622}, L9

\bibitem[\protect\astroncite{{Steinmetz} and {Navarro}}{2002}]{steinnav}
{Steinmetz}, M. and {Navarro}, J.~F.: 2002,
\newblock {\em New Astronomy} {\bf 7}, 155

\bibitem[\protect\astroncite{{Stetson}}{1987}]{stetson}
{Stetson}, P.~B.: 1987,
\newblock {\em PASP} {\bf 99}, 191

\bibitem[\protect\astroncite{{Strauss} et~al.}{2002}]{strauss}
{Strauss}, M.~A., {Weinberg}, D.~H., {Lupton}, R.~H., {Narayanan}, V.~K.,
  {Annis}, J., {Bernardi}, M., {Blanton}, M., {Burles}, S., {Connolly}, A.~J.,
  {Dalcanton}, J., {Doi}, M., {Eisenstein}, D., {Frieman}, J.~A., {Fukugita},
  M., {Gunn}, J.~E., {Ivezi{\'c}}, {\v Z}., {Kent}, S., {Kim}, R.~S.~J.,
  {Knapp}, G.~R., {Kron}, R.~G., {Munn}, J.~A., {Newberg}, H.~J., {Nichol},
  R.~C., {Okamura}, S., {Quinn}, T.~R., {Richmond}, M.~W., {Schlegel}, D.~J.,
  {Shimasaku}, K., {SubbaRao}, M., {Szalay}, A.~S., {Vanden Berk}, D.,
  {Vogeley}, M.~S., {Yanny}, B., {Yasuda}, N., {York}, D.~G., and {Zehavi}, I.:
  2002,
\newblock {\em \aj} {\bf 124}, 1810

\bibitem[\protect\astroncite{{Tinsley}}{1978}]{tinsley}
{Tinsley}, B.~M.: 1978,
\newblock {\em ApJ} {\bf 222}, 14

\bibitem[\protect\astroncite{{Toomre} and {Toomre}}{1972}]{toomres}
{Toomre}, A. and {Toomre}, J.: 1972,
\newblock {\em \apj} {\bf 178}, 623

\bibitem[\protect\astroncite{{Toth} and {Ostriker}}{1992}]{toth}
{Toth}, G. and {Ostriker}, J.~P.: 1992,
\newblock {\em \apj} {\bf 389}, 5

\bibitem[\protect\astroncite{{Trenti} and {Stiavelli}}{2007}]{trenti}
{Trenti}, M. and {Stiavelli}, M.: 2007,
\newblock {\em ArXiv e-prints} 712

\bibitem[\protect\astroncite{{Treu} et~al.}{2005}]{treu}
{Treu}, T., {Ellis}, R.~S., {Liao}, T.~X., {van Dokkum}, P.~G., {Tozzi}, P.,
  {Coil}, A., {Newman}, J., {Cooper}, M.~C., and {Davis}, M.: 2005,
\newblock {\em \apj} {\bf 633}, 174

\bibitem[\protect\astroncite{{Trujillo} and {Aguerri}}{2004}]{moffat}
{Trujillo}, I. and {Aguerri}, J.~A.~L.: 2004,
\newblock {\em \mnras} {\bf 355}, 82

\bibitem[\protect\astroncite{{Trujillo} and {Pohlen}}{2005}]{truj2005}
{Trujillo}, I. and {Pohlen}, M.: 2005,
\newblock {\em \apjl} {\bf 630}, L17

\bibitem[\protect\astroncite{{van den Bosch}}{1998}]{vandenB}
{van den Bosch}, F.~C.: 1998,
\newblock {\em ApJ} {\bf 507}, 601

\bibitem[\protect\astroncite{{van Dokkum} and {Stanford}}{2003}]{vandokkum}
{van Dokkum}, P.~G. and {Stanford}, S.~A.: 2003,
\newblock {\em \apj} {\bf 585}, 78

\bibitem[\protect\astroncite{{Vogt} et~al.}{1996}]{vogt}
{Vogt}, N.~P., {Forbes}, D.~A., {Phillips}, A.~C., {Gronwall}, C., {Faber},
  S.~M., {Illingworth}, G.~D., and {Koo}, D.~C.: 1996,
\newblock {\em ApJL} {\bf 465}, L15+

\bibitem[\protect\astroncite{{Zheng} et~al.}{2007}]{zheng}
{Zheng}, X.~Z., {Bell}, E.~F., {Papovich}, C., {Wolf}, C., {Meisenheimer}, K.,
  {Rix}, H.-W., {Rieke}, G.~H., and {Somerville}, R.: 2007,
\newblock {\em \apjl} {\bf 661}, L41

\end{thebibliography}
\bibliographystyle{astron}

\end{document}